\PassOptionsToPackage{unicode}{hyperref}
\PassOptionsToPackage{hyphens}{url}
\documentclass[
  a4paper,
]{article}
\usepackage{amsmath,amssymb}
\usepackage{iftex}
\ifPDFTeX
  \usepackage[T1]{fontenc}
  \usepackage[utf8]{inputenc}
  \usepackage{textcomp} 
\else 
  \usepackage{unicode-math} 
  \defaultfontfeatures{Scale=MatchLowercase}
  \defaultfontfeatures[\rmfamily]{Ligatures=TeX,Scale=1}
\fi
\usepackage{lmodern}
\ifPDFTeX\else
\fi
\IfFileExists{upquote.sty}{\usepackage{upquote}}{}
\IfFileExists{microtype.sty}{
  \usepackage[]{microtype}
  \UseMicrotypeSet[protrusion]{basicmath} 
}{}
\makeatletter
\@ifundefined{KOMAClassName}{
  \IfFileExists{parskip.sty}{%
    \usepackage{parskip}
  }{
    \setlength{\parindent}{0pt}
    \setlength{\parskip}{6pt plus 2pt minus 1pt}}
}{
  \KOMAoptions{parskip=half}}
\makeatother
\usepackage{xcolor}
\usepackage[top=30mm,left=30mm,right=30mm]{geometry}
\usepackage{longtable,booktabs,array}
\usepackage{calc} 
\usepackage{etoolbox}
\makeatletter
\patchcmd\longtable{\par}{\if@noskipsec\mbox{}\fi\par}{}{}
\makeatother
\IfFileExists{footnotehyper.sty}{\usepackage{footnotehyper}}{\usepackage{footnote}}
\makesavenoteenv{longtable}
\newlength{\cslhangindent}
\newlength{\csllabelwidth}
\newlength{\cslentryspacingunit} 
\newenvironment{CSLReferences}[2] 
 {
  \setlength{\parindent}{0pt}
  \ifodd #1
  \let\oldpar\par
  \def\par{\hangindent=\cslhangindent\oldpar}
  \fi
  \setlength{\parskip}{#2\cslentryspacingunit}
 }%
 {}
\usepackage{calc}

\newcommand{\CSLLeftMargin}[1]{\parbox[t]{\csllabelwidth}{#1}}
\newcommand{\CSLRightInline}[1]{\parbox[t]{\linewidth - \csllabelwidth}{#1}\break}

\usepackage{pgfplots}
\ifLuaTeX
  \usepackage{selnolig}  
\fi
\IfFileExists{bookmark.sty}{\usepackage{bookmark}}{\usepackage{hyperref}}
\IfFileExists{xurl.sty}{\usepackage{xurl}}{} 
\hypersetup{
  pdftitle={Users prefer Jpegli over same-sized libjpeg-turbo or MozJPEG},
  pdfauthor={Martin Bruse, Luca Versari, Zoltan Szabadka, Jyrki Alakuijala},
  hidelinks,
  pdfcreator={LaTeX via pandoc}}

\title{Users prefer Jpegli over same-sized libjpeg-turbo or MozJPEG}
\author{Martin Bruse\footnote{Google Research, Kungsbron 2, 111 22
  Stockholm, Sweden}, Luca Versari\footnote{Google Research,
  Brandschenkestrasse 110, 8002 Zürich, Switzerland}, Zoltan
Szabadka\footnotemark[2], Jyrki Alakuijala\footnotemark[2]}
\date{}

\begin{document}
\maketitle
\begin{abstract}
\noindent We performed pairwise comparisons by human raters of JPEG
images from MozJPEG{[}1{]}, libjpeg-turbo{[}2{]} and our new
Jpegli{[}3{]} encoder. When compressing images at a quality similar to
libjpeg-turbo quality 95, the Jpegli images were 54\% likely to be
preferred over both libjpeg-turbo and MozJPEG images, but used only 2.8
bits per pixel compared to libjpeg-turbo and MozJPEG that used 3.8 and
3.5 bits per pixel respectively. \vskip1em\noindent The raw
ratings{[}4{]} and source images are publicly available for further
analysis and study.
\end{abstract}

\hypertarget{introduction}{%
\subsection{Introduction}\label{introduction}}

We created a new JPEG encoder and decoder, Jpegli, both API-, and
ABI-compatible with libjpeg, MozJPEG, and libjpeg-turbo. The new encoder
uses an adaptive dead-zone quantization technique, and quantization
matrices tuned for combined Butteraugli{[}5{]} and SSIMULACRA{[}6{]}
metrics. We studied which encoder output human evaluators prefer.

The settings were based on the methodology used by CLIC 2024{[}7{]}:
Pairwise comparisons between images, both in standard JPEG format, to
see which one human raters prefer. A publicly available{[}8{]} procedure
was used to compute the Elo{[}9{]} score per method that best explains
the results.

\hypertarget{materials-and-methods}{%
\subsection{Materials and Methods}\label{materials-and-methods}}

The choices of source images, degradation methods, viewing environment,
experimental design, and human raters are described in the following
sections.

\hypertarget{source-images}{%
\subsubsection{Source images}\label{source-images}}

To ensure a meaningful evaluation of compressor quality and output size,
we selected the validation set of the publicly available CID22{[}10{]}
dataset from Cloudinary, consisting of 49 images with known source and
processing. The images were chosen to cover a wide range of contents,
and are composed of a mixture of people, objects, scenery, and graphical
elements.

\hypertarget{image-degradation-method}{%
\subsubsection{Image degradation
method}\label{image-degradation-method}}

We tested three encoders which offer a distortion vs size trade-off via
a single quality parameter and a chroma subsampling setting. We varied
the quality parameter between 55 and 95 for Jpegli and libjpeg-turbo,
and between 70 and 95 for MozJPEG, and vary the chroma subsampling
setting between YUV420 and YUV444 for Jpegli, and between YUV420,
YUV422, and YUV444 for MozJPEG and libjpeg-turbo.

Each combination of encoder, quality parameter, and chroma subsampling
setting defines a degradation method.

The Jpegli degraded images were produced using the Linux shell command
\texttt{cjpegli\ input.png\ output.jpeg\ -\/-quality\ {[}QUALITY{]}\ -\/-chroma\_subsampling={[}444\textbar{}420{]}},
and the MozJPEG- and libjpeg-turbo-degraded images were produced using
the command
\texttt{env\ LD\_LIBRARY\_PATH=\$HOME/{[}mozjpeg\textbar{}libjpeg-turbo{]}/build/\ convert\ input.png\ -quality\ {[}QUALITY{]}\ -sampling-factor\ {[}4:4:4\textbar{}4:2:2\textbar{}4:2:0{]}\ output.jpeg}.

\hypertarget{viewing-environment}{%
\subsubsection{Viewing environment}\label{viewing-environment}}

The human evaluators viewed the images under their own normal viewing
preferences - the experiment controlled neither lighting conditions nor
computer screens or display settings. As such, the ratings produced by
this experiment are expected to align well with in-the-field use of the
JPEG libraries under study.

\hypertarget{experiment-design}{%
\subsubsection{Experiment design}\label{experiment-design}}

We chose a pairwise comparison model, and modeled the probability that a
human evaluator would prefer each method over each other method as a
random variable presented as an Elo score.

We chose pairs of images to compare based on which result would reduce
the uncertainty of the Elo score the most. The human evaluators were
shown the original image along with one of the degraded images, and were
allowed to replace the degraded image with the other as many times as
necessary to make the decision of which degraded image is closest to the
original. The evaluators were forced to make a decision for each
presented pair. This approach leads to methods with difficult to
distinguish images having Elo scores close together.

To ensure evaluator quality, we seeded the images with `golden'
questions where one of the degraded images was the original image and
the other a heavily degraded image, e.g.~a JPEG image with quality 50,
and prevented assigning more questions to evaluators that provided the
wrong answer to multiple `golden' questions.

\hypertarget{subjects}{%
\subsubsection{Subjects}\label{subjects}}

18 raters participated in our experiment. Subjects were recruited from
an internal Google team of human raters specializing in providing
high-quality data to machine learning systems. Subjects were not
specially selected for age, gender, experience, or background.

\hypertarget{results}{%
\subsection{Results}\label{results}}

Each subject generated between 86 and 995 answers during 58 days, with a
mean of 749, a median of 912, and a standard deviation of 340. The
results were analyzed by computing the Elo scores that best explain the
evaluated results, following the CLIC 2024 methodology.

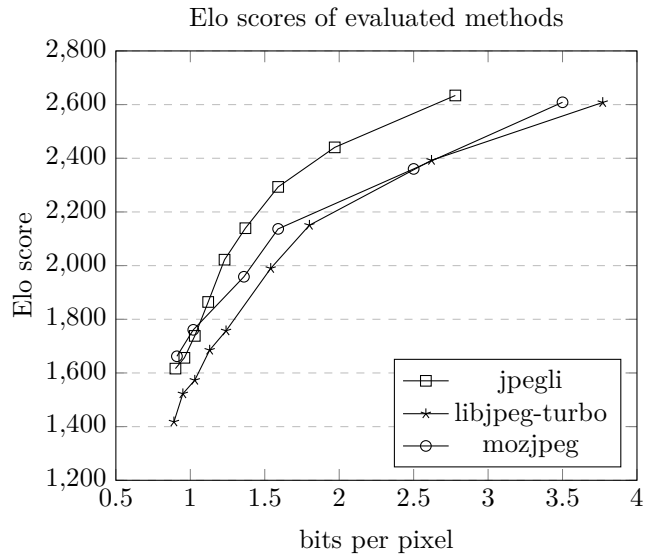
\begin{center} 
\begin{tikzpicture}
\begin{axis}[
    title={Elo scores of evaluated methods},
    xlabel={bits per pixel},
    ylabel={Elo score},
    xmin=0.5, xmax=4.0,
    ymin=1200.0, ymax=2800.0,
    xtick={0.5,1.0,1.5,2.0,2.5,3.0,3.5,4.0},
    ytick={1200,1400,1600,1800,2000,2200,2400,2600,2800},
    legend pos=south east,
    ymajorgrids=true,
    grid style=dashed,
]

\addplot[
    color=black,
    mark=square,
    ]
    coordinates {
    (0.9,1616.22)(0.96,1656.44)(1.03,1738.07)(1.12,1864.52)(1.23,2022.06)(1.37,2139.41)(1.59,2293.27)(1.97,2440.64)(2.78,2634.02)
    };

\addplot[
    color=black,
    mark=star,
    ]
    coordinates {
    (0.89,1417.72)(0.95,1522.56)(1.03,1572.97)(1.13,1685.13)(1.24,1757.05)(1.54,1989.65)(1.8,2150.56)(2.62,2392.53)(3.77,2608.02)
    };

\addplot[
    color=black,
    mark=o,
    ]
    coordinates {
    (0.91,1662.13)(1.02,1760.71)(1.36,1958.47)(1.59,2136.87)(2.5,2360.31)(3.5,2608.61)
    };

\legend{jpegli,libjpeg-turbo,mozjpeg}

\end{axis}
\end{tikzpicture}
\end{center}

Overall, for bitrates above 1 bit per pixel the evaluators were always
more likely to prefer jpegli images to libjpeg-turbo or MozJPEG at
similar bitrates. The computed Elo scores, their 99\% credible
intervals, and mean bits per pixel is listed in Appendix A.

\hypertarget{discussion}{%
\subsection{Discussion}\label{discussion}}

Our headline result is that Jpegli produces JPEGs preferred by human
raters at a lower bitrate using a library that is both API- and
ABI-compatible with libjpeg-turbo and MozJPEG.

To simplify comparison of bitrates at normalized quality levels, we
performed a linear interpolation of the bitrates at each Elo score for
the compared methods. The code used is publicly available{[}11{]}. The
results of this interpolation for the Elo scores of each evaluated
libjpeg-turbo setting is listed in Appendix B.

Our results show clear evaluator preference for Jpegli with bitrates
over 1 bit per pixel, which according to the HTTP Archive's Web Almanac
2022{[}12{]} accounts for 75\% of all images found on the internet.
According to the same source, the median bitrate for JPEG images is 2.1
bits per pixel. According to our interpolation libjpeg-turbo at 2.1 bits
per pixel corresponds to an Elo of approximately 2238, while Jpegli at
Elo 2238 corresponds to a bitrate of approximately 1.5 bits per pixel,
or a reduction of about 28\%.

\hypertarget{conclusion}{%
\subsection{Conclusion}\label{conclusion}}

In a pairwise comparison by 18 human raters of Jpegli, libjpeg-turbo,
and MozJPEG encodings of 49 publicly available images, computed Elo
scores show that at comparable bitrates jpegli encodings are likelier to
be preferred at rates over 1 bit per pixel. We conclude that Jpegli
generates the same or higher quality images at the same or lower
bitrates.

\hypertarget{references}{%
\section{References}\label{references}}

\hypertarget{refs}{}
\begin{CSLReferences}{0}{0}
\leavevmode\vadjust pre{\hypertarget{ref-mozjpeg}{}}%
\CSLLeftMargin{{[}1{]} }%
\CSLRightInline{Mozilla Foundation. {Mozilla JPEG Encoder Project} 2014.
\url{https://github.com/mozilla/mozjpeg}.}

\leavevmode\vadjust pre{\hypertarget{ref-libjpegturbo}{}}%
\CSLLeftMargin{{[}2{]} }%
\CSLRightInline{libjpeg-turbo contributors. {libjpeg-turbo} 2015.
\url{https://libjpeg-turbo.org/}.}

\leavevmode\vadjust pre{\hypertarget{ref-jpegli}{}}%
\CSLLeftMargin{{[}3{]} }%
\CSLRightInline{Jpegli contributors. Jpegli 2022.
\url{https://github.com/libjxl/libjxl/tree/main/lib/jpegli}.}

\leavevmode\vadjust pre{\hypertarget{ref-mucped23answers}{}}%
\CSLLeftMargin{{[}4{]} }%
\CSLRightInline{Versari L, Szabadka Z, Bruse M, Alakuijala J. {mucped23
answers} 2023.
\url{https://github.com/google-research/google-research/blob/master/mucped23/answers.csv}.}

\leavevmode\vadjust pre{\hypertarget{ref-butteraugli}{}}%
\CSLLeftMargin{{[}5{]} }%
\CSLRightInline{Butteraugli contributors. Butteraugli 2016.
\url{https://github.com/google/butteraugli}.}

\leavevmode\vadjust pre{\hypertarget{ref-ssimulacra}{}}%
\CSLLeftMargin{{[}6{]} }%
\CSLRightInline{SSIMULACRA contributors. {SSIMULACRA} 2017.
\url{https://github.com/cloudinary/ssimulacra}.}

\leavevmode\vadjust pre{\hypertarget{ref-clic2024}{}}%
\CSLLeftMargin{{[}7{]} }%
\CSLRightInline{Ballé J, Toderici G, Versari L, Johnston N, Theis L,
Norkin A, et al. {CLIC 2024} 2024. \url{https://compression.cc/}.}

\leavevmode\vadjust pre{\hypertarget{ref-eloratermodel}{}}%
\CSLLeftMargin{{[}8{]} }%
\CSLRightInline{Versari L. {ELO-based pairwise comparison system with
rater modeling} 2024.
\url{https://github.com/google-research/google-research/tree/master/elo/_rater/_model}.}

\leavevmode\vadjust pre{\hypertarget{ref-elo}{}}%
\CSLLeftMargin{{[}9{]} }%
\CSLRightInline{Wikipedia contributors. {Elo rating system} 2024.
\url{https://en.wikipedia.org/wiki/Elo/_rating/_system}.}

\leavevmode\vadjust pre{\hypertarget{ref-cid22}{}}%
\CSLLeftMargin{{[}10{]} }%
\CSLRightInline{Sneyers J, Ben Baruch E, Vaxman Y. {Cloudinary Image
Dataset '22} 2022. \url{https://cloudinary.com/labs/cid22}.}

\leavevmode\vadjust pre{\hypertarget{ref-qualityinterpolation}{}}%
\CSLLeftMargin{{[}11{]} }%
\CSLRightInline{Bruse M. {Quality interpolation Python notebook} 2024.
\url{https://github.com/google-research/google-research/blob/master/mucped23/equivalent/_quality/_interpolation.ipynb}.}

\leavevmode\vadjust pre{\hypertarget{ref-webalmanac}{}}%
\CSLLeftMargin{{[}12{]} }%
\CSLRightInline{{Web Almanac '22, Media} 2022.
\url{https://almanac.httparchive.org/en/2022/media}.}

\end{CSLReferences}

\newpage

\hypertarget{appendix-a}{%
\section{Appendix A}\label{appendix-a}}

\begin{longtable}[]{@{}
  >{\raggedright\arraybackslash}p{(\columnwidth - 8\tabcolsep) * \real{0.4898}}
  >{\raggedright\arraybackslash}p{(\columnwidth - 8\tabcolsep) * \real{0.1429}}
  >{\raggedright\arraybackslash}p{(\columnwidth - 8\tabcolsep) * \real{0.1429}}
  >{\raggedright\arraybackslash}p{(\columnwidth - 8\tabcolsep) * \real{0.1429}}
  >{\raggedright\arraybackslash}p{(\columnwidth - 8\tabcolsep) * \real{0.0816}}@{}}
\toprule\noalign{}
\begin{minipage}[b]{\linewidth}\raggedright
method
\end{minipage} & \begin{minipage}[b]{\linewidth}\raggedright
elo
\end{minipage} & \begin{minipage}[b]{\linewidth}\raggedright
p99Low
\end{minipage} & \begin{minipage}[b]{\linewidth}\raggedright
p99Hi
\end{minipage} & \begin{minipage}[b]{\linewidth}\raggedright
bpp
\end{minipage} \\
\midrule\noalign{}
\endhead
\bottomrule\noalign{}
\endlastfoot
jpegli-q55-yuv444 & 1616.22 & 1570.12 & 1662.31 & 0.9 \\
jpegli-q60-yuv444 & 1656.44 & 1611.1 & 1701.78 & 0.96 \\
jpegli-q65-yuv420 & 1600.83 & 1554.37 & 1647.3 & 0.87 \\
jpegli-q65-yuv444 & 1738.07 & 1695.75 & 1780.39 & 1.03 \\
jpegli-q70-yuv420 & 1692.46 & 1648.48 & 1736.43 & 0.95 \\
jpegli-q70-yuv444 & 1864.52 & 1824.31 & 1904.73 & 1.12 \\
jpegli-q75-yuv420 & 1823.93 & 1782.02 & 1865.83 & 1.05 \\
jpegli-q75-yuv444 & 2022.06 & 1980.59 & 2063.53 & 1.23 \\
jpegli-q80-yuv420 & 1980.54 & 1935.38 & 2025.7 & 1.18 \\
jpegli-q80-yuv444 & 2139.41 & 2096.15 & 2182.67 & 1.37 \\
jpegli-q85-yuv420 & 2135.4 & 2088.14 & 2182.65 & 1.36 \\
jpegli-q85-yuv444 & 2293.27 & 2247.15 & 2339.39 & 1.59 \\
jpegli-q90-yuv420 & 2296.94 & 2247.53 & 2346.35 & 1.7 \\
jpegli-q90-yuv444 & 2440.64 & 2388.39 & 2492.89 & 1.97 \\
jpegli-q95-yuv420 & 2481.99 & 2425.26 & 2538.71 & 2.41 \\
jpegli-q95-yuv444 & 2634.02 & 2568.28 & 2699.77 & 2.78 \\
libjpeg-turbo-q55-yuv420 & 1417.72 & 1359.51 & 1475.93 & 0.89 \\
libjpeg-turbo-q60-yuv420 & 1522.56 & 1472.45 & 1572.67 & 0.95 \\
libjpeg-turbo-q65-yuv420 & 1572.97 & 1525.96 & 1619.99 & 1.03 \\
libjpeg-turbo-q70-yuv420 & 1685.13 & 1642.59 & 1727.67 & 1.13 \\
libjpeg-turbo-q75-yuv420 & 1757.05 & 1714.45 & 1799.65 & 1.24 \\
libjpeg-turbo-q80-yuv422 & 1989.65 & 1947 & 2032.29 & 1.54 \\
libjpeg-turbo-q85-yuv422 & 2150.56 & 2107.5 & 2193.62 & 1.8 \\
libjpeg-turbo-q90-yuv444 & 2392.53 & 2341.05 & 2444.02 & 2.62 \\
libjpeg-turbo-q95-yuv444 & 2608.02 & 2546.63 & 2669.4 & 3.77 \\
mozjpeg-q70-yuv420 & 1662.13 & 1619.22 & 1705.03 & 0.91 \\
mozjpeg-q75-yuv420 & 1760.71 & 1719.52 & 1801.9 & 1.02 \\
mozjpeg-q80-yuv422 & 1958.47 & 1916.88 & 2000.06 & 1.36 \\
mozjpeg-q85-yuv422 & 2136.87 & 2093.91 & 2179.82 & 1.59 \\
mozjpeg-q90-yuv444 & 2360.31 & 2312.13 & 2408.48 & 2.5 \\
mozjpeg-q95-yuv444 & 2608.61 & 2546.69 & 2670.53 & 3.5 \\
\end{longtable}

\newpage

\hypertarget{appendix-b}{%
\section{Appendix B}\label{appendix-b}}

\begin{longtable}[]{@{}
  >{\raggedright\arraybackslash}p{(\columnwidth - 8\tabcolsep) * \real{0.3596}}
  >{\raggedright\arraybackslash}p{(\columnwidth - 8\tabcolsep) * \real{0.0787}}
  >{\raggedright\arraybackslash}p{(\columnwidth - 8\tabcolsep) * \real{0.2360}}
  >{\raggedright\arraybackslash}p{(\columnwidth - 8\tabcolsep) * \real{0.1685}}
  >{\raggedright\arraybackslash}p{(\columnwidth - 8\tabcolsep) * \real{0.1573}}@{}}
\toprule\noalign{}
\begin{minipage}[b]{\linewidth}\raggedright
libjpeg\_turbo\_equiv\_quality
\end{minipage} & \begin{minipage}[b]{\linewidth}\raggedright
elo
\end{minipage} & \begin{minipage}[b]{\linewidth}\raggedright
libjpeg\_turbo\_bitrate
\end{minipage} & \begin{minipage}[b]{\linewidth}\raggedright
mozjpeg\_bitrate
\end{minipage} & \begin{minipage}[b]{\linewidth}\raggedright
jpegli\_bitrate
\end{minipage} \\
\midrule\noalign{}
\endhead
\bottomrule\noalign{}
\endlastfoot
libjpeg-turbo-q70-yuv420 & 1685.13 & 1.13 & 0.94 & 0.99 \\
libjpeg-turbo-q75-yuv420 & 1757.05 & 1.24 & 1.02 & 1.04 \\
libjpeg-turbo-q80-yuv422 & 1989.65 & 1.54 & 1.40 & 1.21 \\
libjpeg-turbo-q85-yuv422 & 2150.56 & 1.80 & 1.65 & 1.39 \\
libjpeg-turbo-q90-yuv444 & 2392.53 & 2.62 & 2.63 & 1.85 \\
libjpeg-turbo-q95-yuv444 & 2608.02 & 3.77 & 3.50 & 2.67 \\
\end{longtable}

\end{document}